\documentclass[pdflatex,sn-basic]{sn-jnl}

\usepackage[english]{babel}
\RequirePackage[l2tabu, orthodox]{nag}

\usepackage{graphicx}
\graphicspath{{figures/}}

\usepackage{booktabs}
\newcommand{\ra}[1]{\renewcommand{\arraystretch}{#1}}

\usepackage[caption=false, font=footnotesize]{subfig}

\usepackage{mathtools}
\DeclarePairedDelimiter{\norm}{\lVert}{\rVert}

\usepackage{xspace}
\makeatletter
\DeclareRobustCommand\onedot{\futurelet\@let@token\@onedot}
\def\@onedot{\ifx\@let@token.\else.\null\fi\xspace}

\def\eg{\emph{e.g}\onedot}

\def\ie{\emph{i.e}\onedot}

\makeatother

\jyear{2023}%

\theoremstyle{thmstyleone}%

\theoremstyle{thmstyletwo}%

\theoremstyle{thmstylethree}%

\usepackage{etoolbox}
\makeatletter
\patchcmd{\ps@headings}
{\hbox to \hsize{\hfill Springer Nature 2021 \LaTeX\ template\hfill}}
{\hbox to \hsize{}}
{}
{}
\patchcmd{\ps@headings}
{\hbox to \hsize{\hfill Springer Nature 2021 \LaTeX\ template\hfill}}
{\hbox to \hsize{}}
{}
{}
\patchcmd{\ps@titlepage}
{\hbox to \hsize{\hfill Springer Nature 2021 \LaTeX\ template\hfill}}
{\hbox to \hsize{}}
{}
{}
\makeatother

\def\makeheadbox{{%
		\hbox to0pt{\vbox{\baselineskip=10dd\hrule\hbox
				to\hsize{\vrule\kern3pt\vbox{\kern3pt
						\hbox{\bfseries Springer Nature:}
						\hbox{\bfseries\itshape Machine Learning for Indoor Localization and Navigation}
						\hbox{This is a post-peer-review, pre-copyedit version of this book chapter.}
						\hbox{The final authenticated version will be available in early 2023.}
						\kern3pt}\hfil\kern3pt\vrule}\hrule}%
			\hss}}}

\begin{document}
	
\makeheadbox

\title[Resource-aware Deep Learning for Wireless Fingerprinting Localization]{Resource-aware Deep Learning for Wireless Fingerprinting Localization}

\author*[1]{\fnm{Gregor} \sur{Cerar}}\email{gregor.cerar@ijs.si}
\equalcont{These authors contributed equally to this work.}

\author[1]{\fnm{Bla\v{z}} \sur{Bertalani\v{c}}}\email{blaz.bertalanic@ijs.si}
\equalcont{These authors contributed equally to this work.}

\author[1]{\fnm{Carolina} \sur{Fortuna}}\email{carolina.fortuna@ijs.si}

\affil*[1]{\orgdiv{Department of Communication Systems}, \orgname{Jo\v{z}ef Stefan Institute}, \orgaddress{\street{Jamova cesta 39}, \city{Ljubljana}, \country{Slovenia}}}

\abstract{%
Location based services, already popular with end users, are now inevitably becoming part of new  wireless infrastructures and emerging business processes. The increasingly popular Deep Learning (DL) artificial intelligence methods perform very well in wireless fingerprinting localization based on extensive indoor radio measurement data. However, with the increasing complexity these methods become computationally very intensive and energy hungry, both for their training and subsequent operation. Considering only mobile users, estimated to exceed 7.4 billion by the end of 2025, and assuming that the networks serving these users will need to perform only one localization per user per hour on average, the machine learning models used for the calculation would need to perform $65 \times 10^{12}$ predictions per year. Add to this equation tens of billions of other connected devices and applications that rely heavily on more frequent location updates, and it becomes apparent that localization will contribute significantly to carbon emissions unless more energy-efficient models are developed and used. In this Chapter, we discuss the latest results and trends in wireless localization and look at paths towards achieving more sustainable AI. We then elaborate on a methodology for computing DL model complexity, energy consumption and carbon footprint and show on a concrete example how to develop a more resource-aware model for fingerprinting. We finally compare relevant works in terms of complexity and training $CO_{2}$ footprint.
}


\maketitle

\section{Introduction}
\label{sec:intro}

The Global Positioning System (GPS) has become the dominant technology for location-based services (LBS) because of its precise, real-time location. Due to line-of-sight (LOS) constraints, GPS is not suitable for indoor use, so alternatives had to be considered. An additional difficulty for LBS indoors is the complexity of the indoor environment, where additional obstacles in the form of furniture, walls, and moving people can alter the propagation of the radio signal in unpredictable ways, resulting in poor reception of LOS and non-line of sight (NLoS) signals. With advances in mobile cellular networks, such as 5G and beyond, communications frequencies are moving toward the millimetre wave (mmWave) band. As \cite{Kanhere_9356512} has shown, mmWave communications combined with steerable high gain MIMO antennas, machine learning (ML), user tracking, and multipath can enable precise smartphone localization in the near future.

Given the widespread use of wireless networks and the associated availability of radio-frequency (RF) measurements, the ML-based algorithms and models offer the greatest guarantee for the development of a high-precision LBS, but at the expense of higher development and deployment costs. Currently, ML is mainly used to improve the existing indoor LBS, by mitigating the effects of range errors due to NLOS and multipath signal propagation. Depending on the technology employed for LBS, NLOS and multipath errors may have different characteristics and advanced high-dimensional patterns suitable for detection by ML. With ML, these properties can be used to either identify and eliminate NLOS/multipath faults or directly attenuate NLOS/multipath effects to directly suppress the error~\citep{Nessa_9264122}. Another common use case for using ML for wireless localization is fingerprinting. During the training phase of a ML model, the collected RF measurements are used to create a fingerprinting database. Later, during the deployment of the trained ML model, real-time RF measurements are fed into the model to predict the exact or estimated location depending on the quality of the model. \cite{savic2015fingerprinting} focused on existing methods for position estimation using classical ML approaches, briefly introducing k-Nearest Neighbour, Support Vector Machine, and Gaussian Process Regression as suitable candidates. But recent studies on fingerprint-based positioning consider larger antenna array and a high number of subcarriers. A large number of antennas and subcarriers inevitably produce rich fingerprint samples and thus a large final dataset that is often hard to leverage by traditional ML approaches. 

Similar to other fields, advances in Deep Learning (DL) combined with larger data sets have enabled the development of more accurate indoor localization algorithms and are considered the most promising approach for next-generation development of LBS~\citep{yan2021extreme}. \cite{arnold2018deep, chin2020intelligent} experimented with fully (\ie, densely) connected layers for fingerprinting. However, it was found that convolutional layers were far more effective than fully connected layers in terms of performance and number of weights. In addition, \cite{widmaier2019towards, bast2020positioning, pollin2020mamimo} examined how CNNs can scale easily with more antennas. The evaluation also shows that more antennas lead to higher accuracy because larger MIMO systems provide more spatial information.

\cite{arnold2019sounding} proposed one of the first simple CNN architectures for fingerprinting. \cite{pollin2020mamimo} proposed a more advanced architecture where they utilized a DenseNet~\citep{huang2017densely} inspired building block for the feature extraction part of the neural network. \cite{bast2020positioning} took a different approach and replaced the convolutional layers for feature extraction with building blocks inspired by ResNet~\citep{he2016residual}. Recent publications by \cite{chin2020intelligent, cerar2021improving, pirnat2022towards} focused on modifications where they emphasize the size of the neural network architecture (\ie, number of weights) in addition to the accuracy of position prediction.

The drawback of the DL method is its complexity, which can negatively impact the energy consumption required for training and predictions. The overall increase in energy consumption with technological advances raises environmental concerns. Therefore, much of the future research will focus on how to reduce the overall environmental impact of various technologies, which includes artificial intelligence (AI) and DL.

\section{Towards Sustainable Green AI}
\label{sec:green-ai}

Recently, the impact of AI technologies has received increased attention from regulators and the public, triggering related research activities in fairness described in \cite{strubell2019energy, strubell2020energy, dhar2020carbon, schwartz2020green}. Furthermore, \cite{schwartz2020green} advocates to increase research effort and investments into Green~AI, which encourages environmental friendliness and inclusiveness.

The cost of building a model from scratch is growing exponentially. \cite{web:ai-compute} estimates 3.4-month doubling time of compute requirements, while for comparison, Moore's Law had a 2-year doubling period. Comparing AlexNet (2012) to AlphaGo Zero (Q4 2017) the compute requirement increased by $300\,000$-times. Similar or even higher increase can be observed in Natural Language Processing (NLP), with BERT and GPT-3, and in Computer Vision (CV) with Vision Transformers (ViT) or Dall$\cdot$E~2, all needing excessive computing power to train compared to state of the art models from a few years ago. \cite{strubell2019energy, strubell2020energy, schwartz2020green} refer to them as Red AI. However, the benefit of these models is undeniable and is helping us move the barriers of what is possible, improve our understanding of AI, and help us see how far we can reach. Unfortunately, these recent models are out of reach for most research communities due to the high cost and time requirements to reproduce, and becoming overly dominant.

One way to reduce the environmental impact of power-hungry AI technology is to increase the proportion of electricity from clean energy sources such as wind, solar, hydro, and nuclear. However, striving only for net-zero energy neutrality is not sustainable and neglects other issues, such as produced heat, noise, and landscape interference/impact. Therefore, it must be complemented by other efforts to optimize energy consumption relative to the performance of existing and emerging technologies such as optimizing AI/DL algorithms and models.

\cite{strubell2019energy, strubell2020energy, garcia2019estimation} studied and estimated the energy consumption of several well-known DL architectures. They conclude that the increasing complexity of the models, manifested in the number of configurable weights, affects their performance and energy consumption. However, what they showed was that more parameters do not always equate to better model performance. This was first observed with the invention of convolutional neural networks (CNN) by \cite{lecun1995convolutional}. CNNs revolutionized the field of CV by significantly improving performance of CV tasks, while reducing the number of adjustable parameters, while also making CNNs easier to scale (\ie, deeper NNs). Subsequently this led to a significant reduction in the required mathematical operations in larger DL architectures.

For more sustainable AI, the architecture of a DL network needs to be optimized for energy efficiency, ideally, done in a way that would not affect the model's performance. There are several ways to achieve that. One option is to propose a novel architecture, building block, or layer to replace an existing one for better efficiency, similar to how CNNs and Transformers revolutionized DL tasks. Another is to use specialized hardware to perform specific tasks more effectively, which may benefit from reduced precision or specialized data types. Final option is to optimize existing DL architecture with the intention of reducing the required mathematical operations.

\subsection{Carbon Footprint of AI models}

Recently researchers started showing more interest in analyzing the carbon footprint of DL algorithms. \cite{gigi2020carbon} analyzed the carbon footprint of both Fully connected and Convolutional layers and concluded that the models with the fewest parameters showed the best relation between the performance and carbon footprint. Furthermore, since the power-hungry nature of DL algorithms is well known, \cite{verhelst2017embedded} discussed the possibility of using different techniques for optimization of hardware for DL, especially for embedded devices.  

\cite{garcia2019estimation} published a survey on the energy consumption of a large variety of models. They presented a set of techniques for model power estimation at the hardware and software level and, at the same time, debated existing approaches for energy consumption estimation. They showed that the number of weights do not directly correlate to energy consumption. They proposed calculating the number of floating-point operations (FLOPs) or multiply-accumulate operations (MACs), which more accurately correlate to the energy consumption of DL models. \cite{jurj2020environmentally} did similar work, where they proposed similar metrics as \cite{garcia2019estimation}, that account for the trade-off between model performance and energy consumption.

\section{Methodology for Calculating Model Complexity, Energy Consumption and Carbon Footprint}
To calculate models potential energy consumption and carbon footprint, model complexity has to be calculated. Model complexity is provided in FLOPs. Since model complexity has to be calculated for each layer separately, the insights from ~\citep{Verhelst2017}\footnote{https://cs231n.github.io/convolutional-networks/\#conv} can be used as a starting point for calculations. In this section we first explain the rationale of computing complexity of the various layers, of entire networks, followed by hardware architecture specific computational complexity. We also present theoretical formulations for computing energy and carbon footprint consumption. We also make available a per-layer online calculator\footnote{https://sensorlab.ijs.si/apps/ccwebapp/}, developed based on the findings from \cite{pirnat2022towards} that may speed up assessing the corresponding computational complexity.     

\subsection{Fully Connected Layer}

\begin{figure}[htbp]
  \centering
  \includegraphics[width=0.6\linewidth]{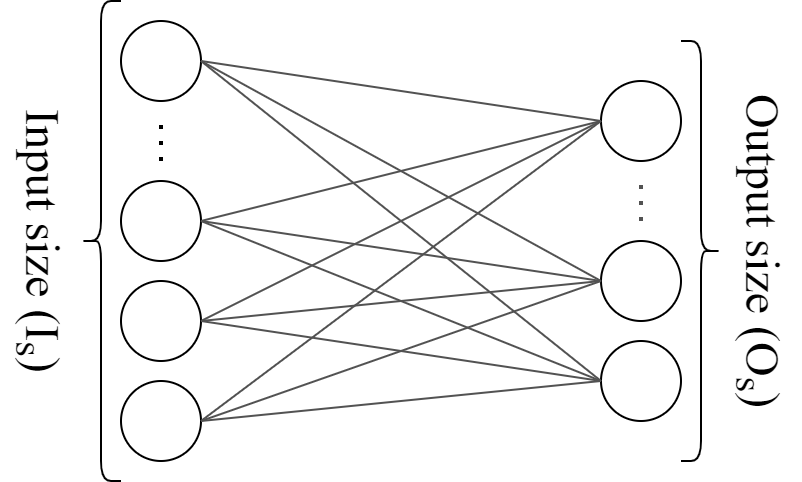}
  \caption{Dense layer input and output layer.}
  \label{fig:calc:dense}
\end{figure}

A Fully Connected or Dense (FC) layer performs multiplication accumulation operations (MAC). The number of operations depends on both the input size $I_\text{s}$ and the output size $O_\text{s}$ as shown in  Fig.\ref{fig:calc:dense}. The FLOPs are calculated according to Eq.~\ref{eq:FLOPSFC}. The total number of FLOPs is equal to the product of the number of input neurons and the number of output neurons multiplied by two, because one MAC operation costs two FLOPs. Also, if biases are used, the size of the output must be added to the results of the first product.
\begin{equation}
\label{eq:FLOPSFC}
F_\text{fc} = 2\,I_\text{s}O_\text{s} + O_\text{s}
\end{equation}

\subsection{Convolutional Layer}

\begin{figure}[htbp]
  \centering
  \includegraphics[width=0.6\linewidth]{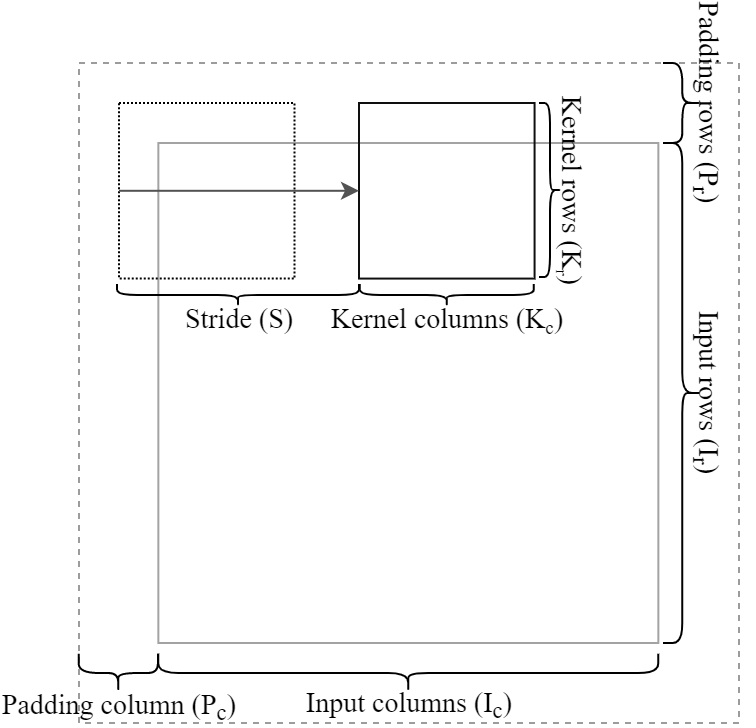}
  \caption{Kernel sliding and padding of the input sample.}
  \label{fig:calc:cnnKernel}
\end{figure}

\begin{figure}[htbp]
  \centering
  \includegraphics[width=0.4\linewidth]{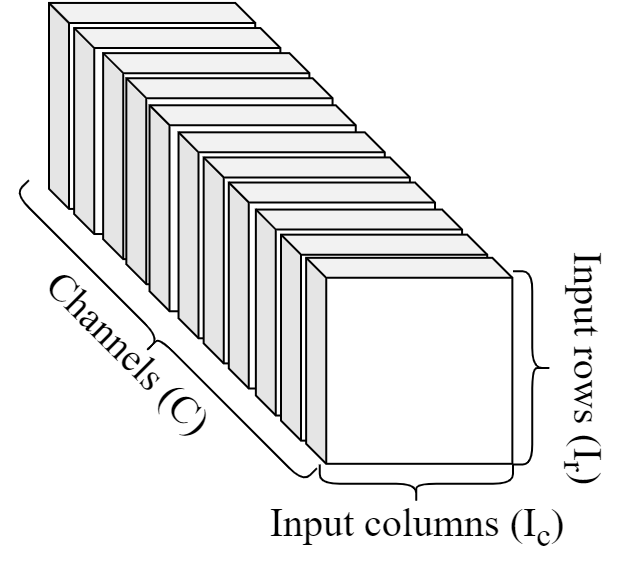}
  \caption{CNN input tensor.}
  \label{fig:calc:cnnInput}
\end{figure}

FLOPs per filter or kernel ($F_{pf}$) of a layer in a convolutional neural network are computed according to  Eq.~\ref{eqn:FlopsPerFilter}, where $K_\text{r} \times K_\text{c}$ represents the size of a set of filters or kernels present in the convolutional layer as presented in Fig.~\ref{fig:calc:cnnKernel}. These kernels are then applied to the input tensor of size $I_\text{r} \times I_\text{c} \times C$, as seen in Fig. \ref{fig:calc:cnnInput}, with a stride over rows $S_r$ and columns $S_c$ as per Fig. \ref{fig:calc:cnnInput}. Finally, $P_r$ and $P_c$ represent the padding used around the input tensor depicted in Fig. \ref{fig:calc:cnnInput} for both rows and columns.

\begin{equation}
\label{eqn:FlopsPerFilter}
F_\text{pf} = (\frac{I_\text{r} - K_\text{r} + 2P_\text{r}}{S_\text{r}} + 1)(\frac{I_\text{c} - K_\text{c} + 2P_\text{c}}{S_\text{c}} + 1) (CK_\text{r}K_\text{c} + 1)
\end{equation}

The term in the first parenthesis gives the height of the output tensor, while the second parenthesis gives the width of the output tensor. The last bracket indicates the depth of the input tensor and the bias.

\begin{equation}
\label{eqn:FlopsAllFilter}
F_\text{c}=F_\text{pf} \times N_\text{f}
\end{equation}

The final number of FLOPs per layer is given in Eq.~\ref{eqn:FlopsAllFilter} and is equal to the number of FLOPs per filter from Eq.~\ref{eqn:FlopsPerFilter} times the number of filters/kernels ($N_\text{f}$). In the case of the ReLU activation function, additional multiplication and comparison is required to calculate the total number of FLOPs in the layer for a training epoch, resulting in Eq.~\ref{eqn:FlopsLayerRelu}:

\begin{equation}
\label{eqn:FlopsLayerRelu}
F_\text{c} = (F_\text{pf} + (CK_\text{r}K_\text{c} + 1))N_\text{f} .
\end{equation}

\subsection{Pooling Layer} 
The use of the pooling layer is common in the domain of DL, as it is responsible for downsampling the input tensor data. Since no padding is performed in pooling, the computation of FLOPs is much simpler compared to the computation for CNN. The pooling layer is responsible for downsampling the height and width of the input tensor. The number of FLOPs per pooling layer $F_\text{p}$ is given by:
\begin{equation}
\label{eqn:FlopsPerPool}
F_\text{p} = (\frac{I_\text{r} - K_\text{r}}{S_\text{r}} +1)(\frac{I_\text{c} - K_\text{c}}{S_\text{c}} + 1) (CK_\text{r}K_\text{c} + 1)
\end{equation}

\subsection{Total Number of FLOPs per Neural Network} 
\label{sec:num_flops}
The number of operations per pass of the input tensor in a DL architecture depends on the number and types of layers $L$. Thus, the final accumulation of FLOPs can be calculated as the sum of FLOPs over all layers used in the DL architecture:
\begin{equation}
\label{eqn:ModelFlops}
M_\textrm{FLOPs} = \sum_{l=1}^{L} F_{l} ,
\end{equation}

where $F_{l}$ refers to the $l\textsuperscript{th}$ layer of the architecture, which for some existing architectures proposed for localization, such as \cite{pirnat2022towards}, can be either $F_{fc}$ from Eq.~\ref{eq:FLOPSFC}, $F_{c}$ from Eq.~\ref{eqn:FlopsLayerRelu}, or $F_{p}$ from Eq.~\ref{eqn:FlopsPerPool}. The final number of FLOPs can then be used to estimate the complexity of the proposed architecture, while directly correlating with the amount of energy required for the training and production phases of the DL model.

Assume an example network having three layers $N_{L=3}$, one of each mentioned so far in this chapter, with the following params: 
\begin{itemize}
    \item An input image of size $100\times 100\times 3$
    \item Convolutional layer with 1 kernel of size  $3\times 3$, with stride 1 and padding of 1, so that the output number of rows and columns stay the same as the input
    \item Pooling layer with kernel of size  $2\times 2$, with stride 2
    \item Fully connected layer with 4 output neurons
\end{itemize}
Then the MFLOPs for this network would be 0.312.  

\subsection{Theoretical Computational Performance}

In computing, one way to measure peak theoretical computing performance is floating-point operations per second (FLOPS)\footnote{https://en.wikichip.org/wiki/flops}. If we simplify to a system with only one CPU, FLOPS are calculated according to Eq.~\ref{eq:peak-flops}. The number of FLOPS depends on floating-point operations per cycle per core (FLOPs), the number of cycles per second on a processor (correlates with core frequency), and CPU cores. Notice the distinction between FLOPS, the number of floating-point operations per second, and FLOPs, the number of operations per cycle per core.

\begin{equation}
\label{eq:peak-flops}
\textrm{FLOPS} = %
    \frac{\textrm{FLOPs}}{\textrm{cycle}}%
    \times%
    \frac{\textrm{cycles}}{\textrm{second}}%
    \times%
    \textrm{cores}
\end{equation}

The number of FLOPs per cycle varies with architecture, and it also depends on the data type used. Therefore, FLOPs are listed per relevant data type natively supported by the architecture. Data types are, for instance, double-precision (FP64), single-precision (FP32), half-precision (FP16), Brain Float-point (BF16), which was developed with DL in mind, 8-bit integer (INT8), and 1-bit integer (\ie bit-array) (INT1). Their native support varies with architecture. On most architectures, reducing precision can provide a significant speed bump for the training and inference process of the DL model. However, with reduced precision, model accuracy will decrease, or it may even fail to converge.

\begin{table}[htbp]
\ra{1.3}
\centering
\caption{FLOPS per core per cycle.}
\label{tab:FLOPs}
\begin{tabular}{lccl}
    \hline
    \multirow{2}{*}{\bfseries Architecture} & \multicolumn{2}{c}{\bfseries FLOPs} & \multirow{2}{*}{\bfseries Misc} \\\cmidrule(lr){2-3}
    {} & \bfseries FP32 & \bfseries FP16 & {} \\
    \midrule
    \multicolumn{4}{c}{\bfseries CPU} \\
    
    ARM Cortex-A72 & 8 & 8 (as FP32) & NEON SIMD \\
    Intel Skylake & 32 & 32 (as FP32) & AVX2, FMA (256-bit) \\
    AMD Zen 2 \& 3 & 32 & 32 (as FP32) & AVX2, FMA (256-bit) \\
    Intel Ice Lake & 64 & 64 (as FP32) & AVX512, FMA (512-bit) \\
    
    \hline
    \multicolumn{4}{c}{\bfseries GPU} \\
    
    Nvidia Pascal, Turing  & 2 (FP32) + 2 (INT32) & 16 & - \\
    Nvidia Ampere & 2 (FP32) + 2 (INT32) & 32 & - \\
    
    \hline
\end{tabular}
\end{table}

As presented in Table~\ref{tab:FLOPs}, current mainstream CPU architectures (\eg Zen, Skylake) take advantage of specialized instructions, such as Advanced Vector eXtension (AVX) and Fused-Multiply-Add (FMA) extensions, to achieve 32 FLOPs on MAC operations. Newer architectures (\eg~Ice Lake) and selected server-grade parts come with support for the AVX-512 instruction set that further increases FLOPs. Furthermore, future CPU architectures are speculated to have an Advanced Matrix eXtensions (AMX) instruction set that will introduce lower precision data types, such as INT8 and BF16, and enable operations over whole matrices.

As opposed to general-purpose CPUs, accelerators, such as GPUs and TPUs, have more freedom regarding implementation constraints, backward incompatibility, and features, which can be added faster. While not so sophisticated as CPUs regarding FLOPs, their strength comes in an enormous amount of parallel cores and memory bandwidth. For comparison, server-grade AMD EPYC (Milan) CPUs come with up to 128 CPU threads, a 512-bit memory bus, and up to 280\,W power draw. At the same time, Nvidia A100 comes with an astounding 6912 CUDA cores, specialized memory with a dedicated 5120-bit memory bus, and 350\,W power draw. Overall, such accelerators are significantly faster in the training and inference process.

\subsection{Theoretical Energy Consumption}
\label{sec:TEC}

The training process can be a very tedious process in the life cycle of developing a DL model. Training involves sequential forward and backward propagation through the number of layers of the architecture. During forward propagation, the loss is calculated based on the initialized weights and biases that are later used in the optimization process of the model. Optimization occurs during the backward propagation process where gradients are calculated so that weights and biases can be updated with the goal of minimizing loss during forward propagation. The training process is performed in epochs or training cycles, where for each epoch all available training samples are fed into the network and forward and backward propagation is computed for each epoch. Depending on the number of epochs, the training process can be very energy and time consuming. On the other hand, predictions require only one forward pass through the network.

During forward propagation, the network calculates the loss based on the initialized weights. During backward propagation, it updates the weights and biases based on the calculated gradients against loss. Training is performed in epochs, where an epoch includes a forward and then a backward movement through all available training samples. Prediction, on the other hand, requires only one forward pass through the network.

To calculate the theoretical energy consumption in Joules during the training process, we first calculate the theoretical energy consumption for forward propagation of a sample ($E_{fp}$), which is shown in  Eq.~\ref{eqn:Forward}. Here, the total number of FLOPs in the network ($M_\textrm{FLOPs}$) is multiplied by the number of training data ($training_{samples}$) and the number of \textit{epochs}. Finally, the FLOPs required for the entire training process for forward propagation are divided by the theoretical $GPU_\textrm{performance}$, which is measured in FLOPS/Watt. The theoretical $GPU_\textrm{performance}$ is provided by the device manufacturer. Although Eq.~\ref{eqn:Forward} is presented for the training process on the Graphical Processing Unit (GPU), it can be easily replaced by any other processing device such as Central Processing Unit (CPU), Tensor Processing Unit (TPU) or similar. 

\begin{equation}
E_{fp} = \frac{M_\textrm{FLOPs} }{GPU_\textrm{performance}}(training_\textrm{samples} \times \textrm{epochs})
\label{eqn:Forward}
\end{equation}

Calculating the theoretical energy consumption for backward propagation ($E_{bp}$) is more difficult, but \cite{devarakonda2017adabatch} has shown that backward propagation for ResNet20 takes about twice as long to calculate as forward propagation. Therefore, we can estimate that $E_{bp}$ is:

\begin{equation}
E_{bp} = 2 \times E_{fp}
\label{eqn:backpropagation}
\end{equation}

Under this assumption, the total theoretical energy required for a training process $E_{T}$ can be estimated as follows:

\begin{equation}
E_\textrm{training} = E_{fp} + E_{bp} = 3 \times E_{fp}
\label{eqn:Training}
\end{equation}

Once the trained model goes into production, the theoretical energy consumption for a prediction is equal to the energy required for a forward pass, where input is the number of input samples for the prediction:

\begin{equation}
E_{prediction}=M_\textrm{FLOPs}/{GPU_\textrm{performance}} \times \textrm{input}
\label{eqn:Training}
\end{equation}

For the example, the network with three layers $N_{L=3}$ used as a guiding example in Section\ref{sec:num_flops}, would be trained with 10000 training samples for 100 epochs on a Nvidia A100 GPU with 445.7 GFLOPS/W. For training the model would consume approximately 0.7 W, while a single prediction would consume 70 $\mu$W.

\subsection{Calculating Theoretical Carbon Footprint}

The calculation of the carbon footprint is different for different areas and countries. It depends on the energy sources used to generate electricity. Countries that rely more on clean energy sources such as wind and solar also produce a smaller amount of carbon per kilowatt-hour than countries that generate more electricity from coal. For example, the U.S. West Coast produces an estimated 250\,g of CO$_2$ equivalent per kilowatt-hour ($CO_{2}eq$)\footnote{https://electricitymap.org/}.

To calculate the carbon footprint of our model, we must first calculate the theoretical energy consumption for training according to the equations in Section~\ref{sec:TEC}. Since the equations presented give the theoretical energy consumption in joules, this must be converted to kilowatt-hours, whereupon we can calculate the carbon footprint as follows:
\begin{equation}
\textrm{carbon footprint} = E_\textrm{training}[kWh] \times CO_{2}eq [g]
\label{eqn:co2}
\end{equation}

The carbon footprint can be calculated for both training and predictions. Although the carbon footprint for training is much higher than for a single prediction, when the model is used in production, the carbon footprint increases linearly with the number of predictions and may exceed the carbon footprint of training in the long run.

For the example network with three layers $N_{L=3}$ used as a guiding example in Section\ref{sec:num_flops}, the estimated carbon footprint according to Eq. \ref{eqn:co2} would be 0.175 g. 

\section{On Designing the PirnatEco Model for Localization}

A relatively recent localization challenge \cite{web:ctw2019dataset} motivated the research community, including our group \cite{cerar2021improving}, to further investigate into localization performance improvements. As LBS services will be central to future cellular systems where mobile users are estimated to exceed 7.4 billion by the end of 2025. Assuming that the networks serving these users will need to perform only one localization per user per hour on average, the machine learning models used for the calculation would need to perform $65 \times 10^{12}$ predictions per year. Add to this  tens of billions of other connected devices and applications that rely heavily on more frequent location updates, and it becomes apparent that localization will contribute significantly to carbon emissions unless more energy-efficient models are developed and used.

Therefore we attempted to explore how to take an existing DL architecture and adapt it in order to perform comparably to the state of the art while significantly reducing the computational complexity and carbon footprint. Our findings, including the resulting PirnatEco architecture, were published in \cite{pirnat2022towards} while in this section we elaborate on the specifics of the data made available in the respective challenge and the design decisions made for developing PirnatEco.  

\subsection{CWT 2019 dataset}
\label{sub:ctw2019}

\begin{figure}[htbp]
  \centering
  \includegraphics[width=0.5\linewidth]{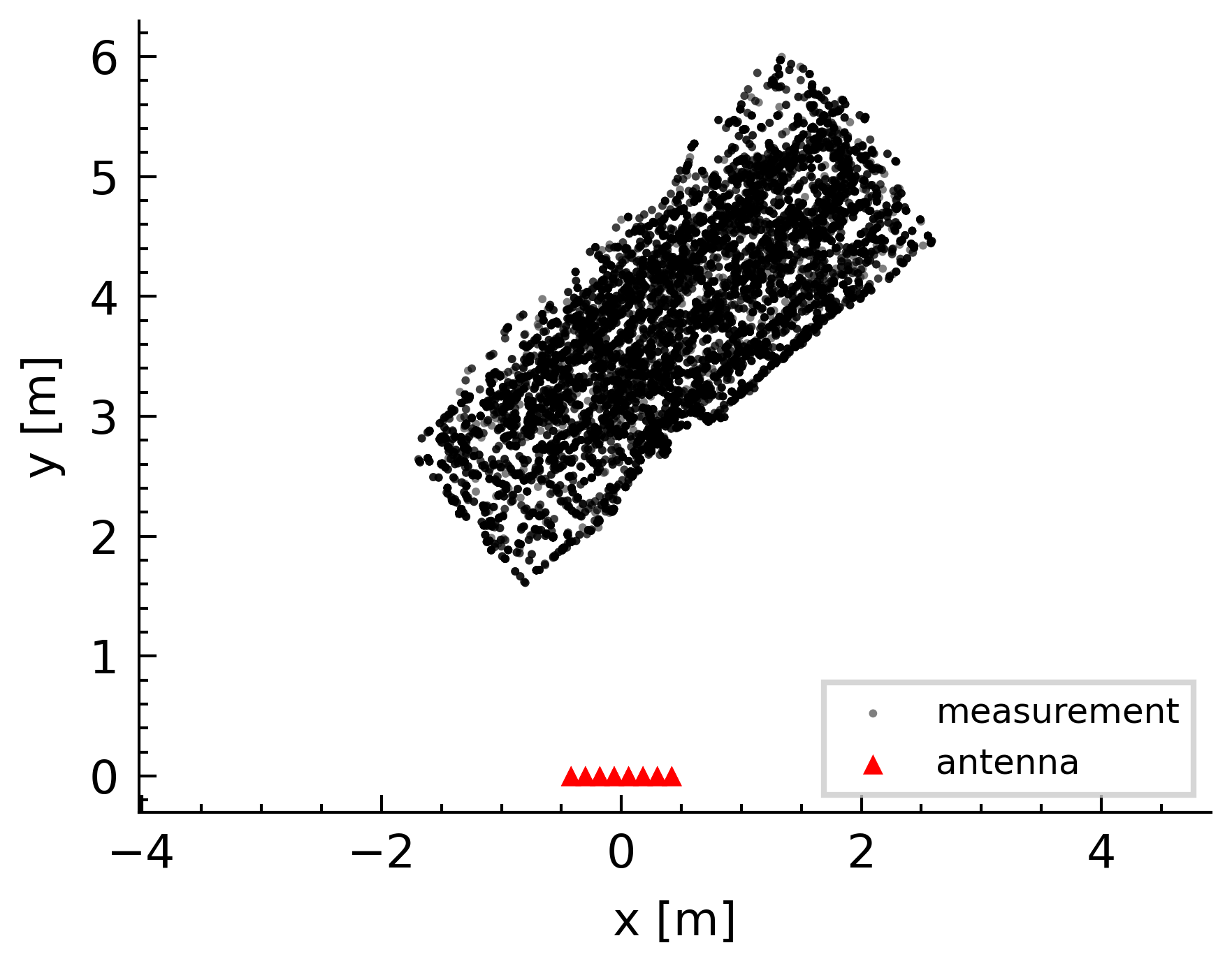}
  \caption{Top-down view on 17\,486 samples and antenna array orientation.}
  \label{fig:ctw2019:samples}
\end{figure}

The dataset was acquired by a massive MIMO channel sounder~\cite{arnold2019sounding} in a setup visually depicted in~\cite{web:ctw2019dataset}. The CSI was measured between a moving transmitter and $8\times2$ antenna array with $\frac{\lambda}{2}$ spacing. The transmitter implemented on SDR was placed on a vacuum cleaner robot. The robot drove in a random path on approximately $4 \times 2$\,m$^2$ size table. The transmitted signal consisted of OFDM pilots with a bandwidth of 20\,MHz and 1024 subcarriers at the central frequency 1.25\,GHz. 100 sub-carriers were used as guard bands, 50 on each side of the frequency band. Figure~\ref{fig:ctw2019:samples} depicts the top-down view of the antenna orientation and positions of 17\,486 CSI samples available in the dataset.

Each raw data sample consists of three components. The first component is channel response tensor $\mathbf{H}$ of shape $(16,\,924,\,2)$. The data thus contain complex (i.e. real and imaginary parts) channel response values of 16 antennas and 924 subcarriers on each. The second data component is $\mathbf{SNR}$ (signal to noise ratio). It is of shape $(16,\,1)$, one value per antenna. Finally, the third data component is a target value given as relative Cartesian X-Y-Z values.

\begin{figure}[htbp]
\centering
\subfloat[closest sample\label{fig:ctw:closest}]{%
    \includegraphics[width=0.49\linewidth]{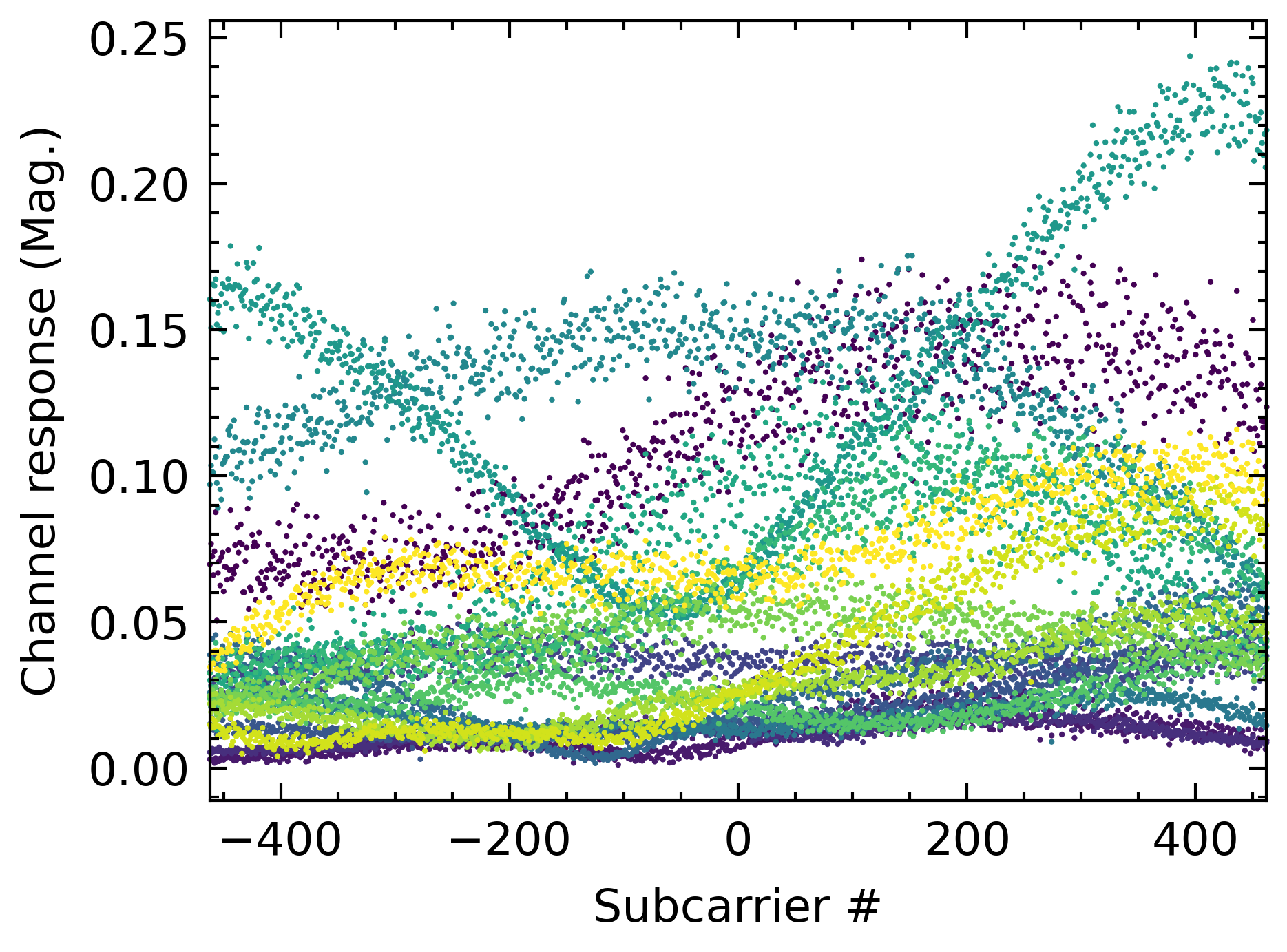}}
\hfill
\subfloat[farthest sample\label{fig:ctw:farthest}]{%
    \includegraphics[width=0.49\linewidth]{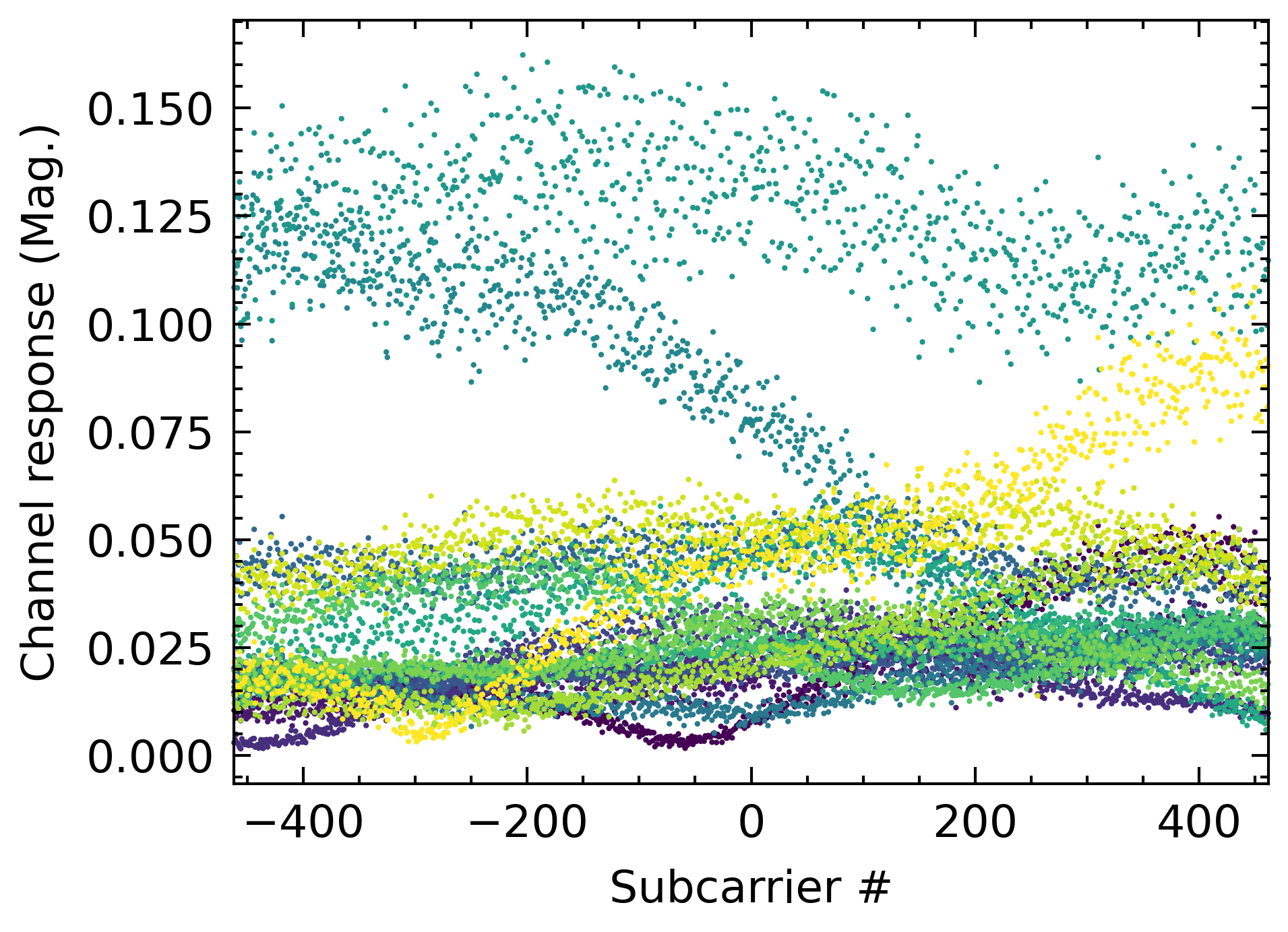}}
 \caption{The magnitude of channel response of closest and farthest sample.}
\label{fig:ctw:csi}
\end{figure}

In Figure~\ref{fig:ctw:csi}, we depicted the absolute channel response $\lVert\mathbf{H}\rVert$ for the closest (Fig.~\ref{fig:ctw:closest}) and the farthest (Fig.~\ref{fig:ctw:farthest}) samples from the antenna array. In both figures, we see 16 ``noisy'' curves distinct by color. Each curve represents channel response on an individual antenna from lowest (left) to highest (right) of 924 subcarriers (\ie, frequencies). If we compare the examples in Fig.~\ref{fig:ctw:closest} and Fig.~\ref{fig:ctw:farthest}, we notice a significant difference in the channel response curves. The difference arises from signal propagation characteristics, where magnitude and phase are dependent on distance and time traveled. The difference can be seen in the dips and rises patterns in the channel response. Moreover, a combination of channel response patterns is a unique ``fingerprint'' for each location, enabling localization.

\subsection{DL Architecture Adaptation}
In DL architectures, one way to optimize the use of energy is to reduce the size of the filters, also referred to as kernels, that represent matrices used to extract features from the input. In these filters, we can adjusts the amount of movement over the image by a stride. Another way is to adjust pools, which represent layers that resize the output of a filter and thus reduce the number of parameters passed to subsequent layers, making a model lighter and faster.
\begin{figure*}[htbp]
    \centering
    \includegraphics[width=\linewidth]{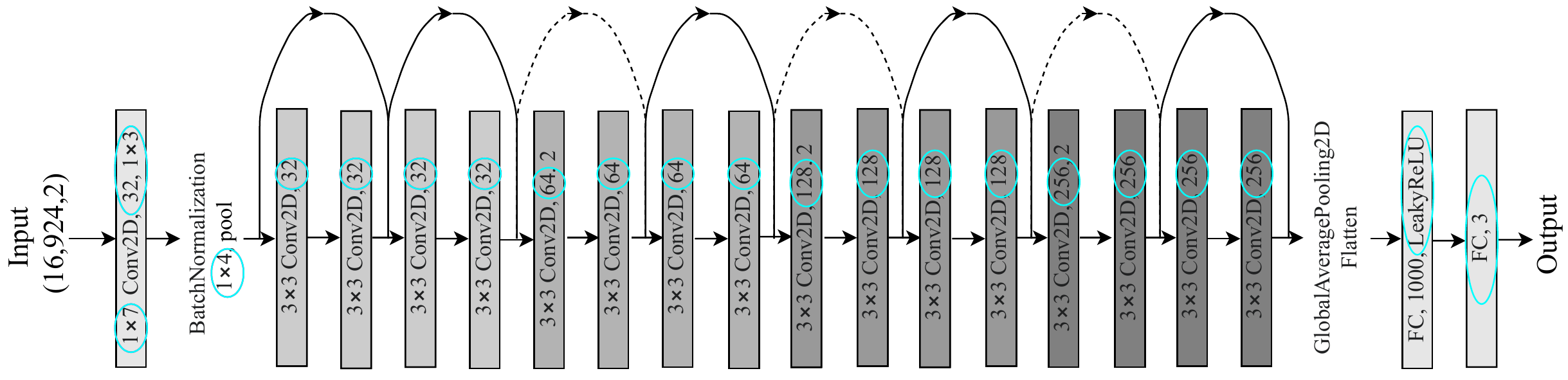}
    \caption{The PirnatEco architecture adapted from ResNet18 with differences marked with cyan circles.}
    \label{fig:Pirnat_1G}
\end{figure*}

In our approach \cite{pirnat2022towards}, we chose ResNet18 because it is the least complex ResNet DL model and is more adaptable to less complex types of images constructed from time series, as is the case with localization. In Figure~\ref{fig:Pirnat_1G}, each layer is visible and explained with its kernel size, type, number of nodes and in some cases stride and activation function. The cyan circles mark the differences with ResNet18. Unlike ResNet18, in PirnatEco the first layer is a convolutional 2D layer (Conv2D) with a kernel size of $1\times7$ and a stride of $1\times3$, followed by a batch normalization and  pooling layer with a pool size of $1\times4$. These kernels and pools are designed to move across the subcarriers of a single antenna as in the CWT challenge dataset, which is different from the square kernels and pools in ResNet18. 

\begin{figure}[htb]
    \centering
    \includegraphics[width=0.8\linewidth]{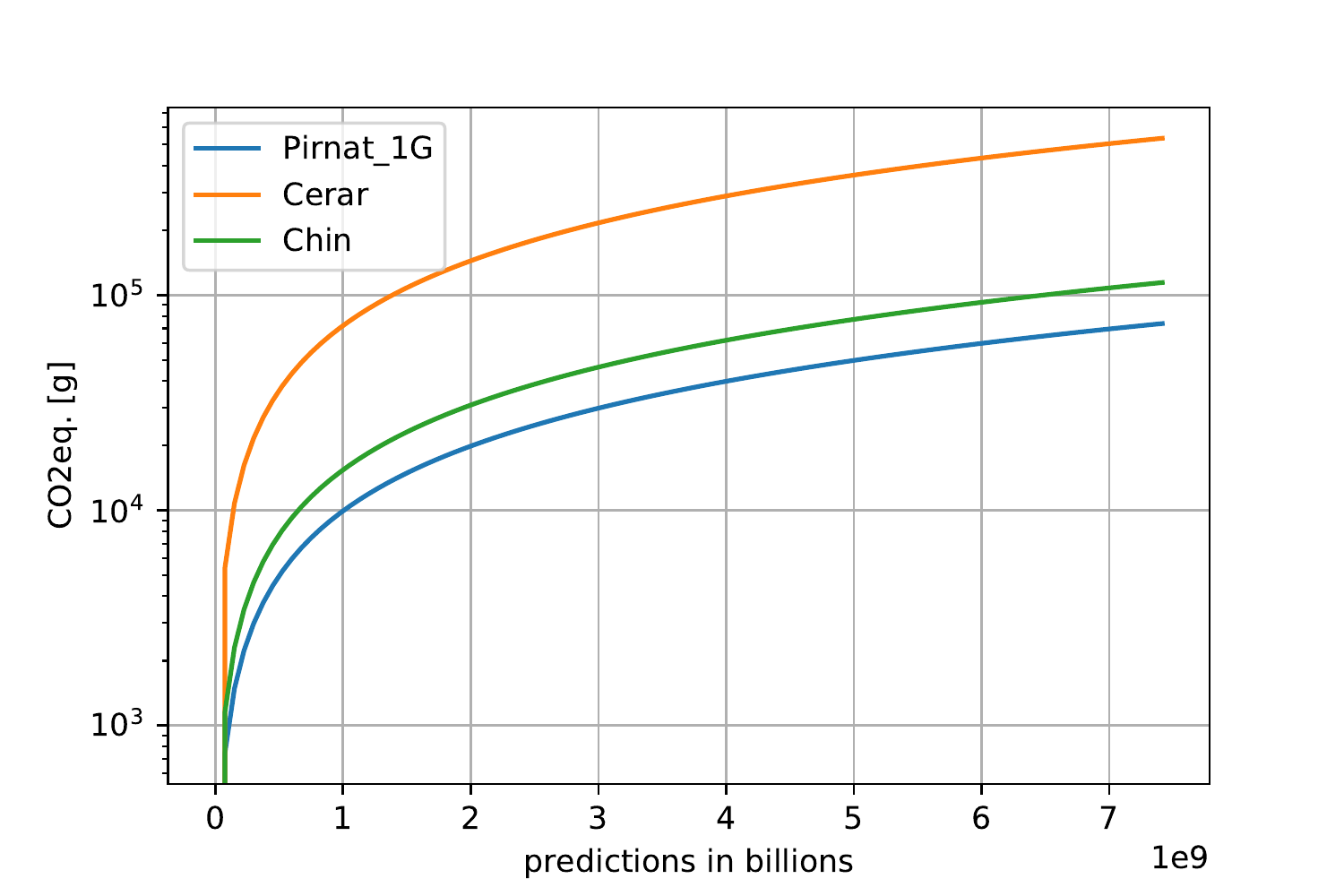}
    \caption{Carbon footprint vs predictions made in logarithmic scale.}
    \label{fig:CTPL}
\end{figure}

Next, we adapted ResNet blocks with reduced number of weights, where the number of nodes doubles every four layers from 32 to 256, unlike ResNet18 which starts with 64. The kernel size in the blocks is $3\times3$, similar to ResNet18. Finally, PirnatEco uses LeakyReLU activation with a parameter alpha set to 10\textsuperscript{-3} at the fully connected (FC) layer with 1000 nodes, unlike ResNet18 which uses ReLU. 

As shown in \cite{pirnat2022towards} and in  Figure~\ref{fig:CTPL}, the CO$_2$ emissions as a function of the number of location predictions of PirnatEco improves on the closest state of the art. The final number in the graph shows CO$_2$ emissions produced if we made only one prediction for each mobile user in 2025 when the estimate number of mobile users is supposed to exceed 7.4 billion.

\section{Performance and Resource Consumption of DL Architectures for Localization}
To estimate the energy efficiency and carbon footprint of the proposed state of the art DL architectures for wireless fingerprinting, they were all tested in controlled settings. All DL architectures were trained on the same CTW2019~\citep{web:ctw2019dataset} dataset and were trained and validated on the same machine with Nvidia T4 graphics card. All models were trained for 200 epochs.

\subsection{Evaluation Metrics}
For evaluation, we found the two most common metrics in the literature. The most commonly used metric is the root mean square error (RMSE)~(\ref{eq:rmse}), which penalizes significant errors more.

\begin{equation}\label{eq:rmse}
  \textrm{RMSE} = \sqrt{ \mathbb{E}\left[ \norm{p - \hat{p}}_2^2 \right]}.
\end{equation}

The second most common metric is normalized mean distance error (NMDE)~(\ref{eq:nmde}). Because of the normalization, errors at samples farther away from the antenna are less penalized. Locations farther from the antenna array are also harder to estimate.

\begin{equation}\label{eq:nmde}
  \textrm{NMDE} = \mathbb{E}\left[\frac{\norm{p - \hat{p}}_2}{\norm{p}_2}\right].
\end{equation}

For the evaluation, the dataset contains the relative transmitter's location ($p = p(x,y,z)$) given in Cartesian coordinates that correspond to the distance from the tachometer. The ultimate goal is to predict the transmitter's location as accurately as possible.

\subsection{Energy consumption and carbon footprint}

\begin{table}[htbp]
\ra{1.3}
\caption{Evaluation of NN for localization in number of weights and RMSE \citep{pirnat2022towards}}
\label{tab:nn:performance}
\centering
\begin{tabular}{llrr}
    \toprule
    \bfseries Approach
    & \bfseries Type
    & \bfseries Weights $[10^6]$
    & \bfseries RMSE $[m]$
    \\\midrule
    
    Dummy (linear) & Linear & $<$0.1 & 0.724 \\
    \cite{bast2020positioning} & CNN & 0.4 & 0.722 \\
    \cite{arnold2018deep} & FC & 32.3 & 0.570 \\
    \cite{chin2020intelligent} & FC & 123.6 & 0.563 \\
    \cite{arnold2018deep} & CNN & 7.6 & 0.315 \\
    \cite{cerar2021improving} CNN4 & CNN & 5.3 & 0.122 \\
    \cite{cerar2021improving} CNN4R & CNN & 10.8 & 0.113 \\
    \cite{pirnat2022towards} PirnatEco & CNN & 3.1 & 0.109 \\
    \cite{cerar2021improving} CNN4S & CNN & 16.3 & 0.108 \\
    \cite{chin2020intelligent} & CNN & 13.7 & 0.100 \\
\bottomrule
\end{tabular}
\end{table}

In Table~\ref{tab:nn:performance}, we see the performance results of proposed NN architectures for fingerprinting tested on CTW2019 data, which are sorted by descending root mean square error (RMSE). As a baseline, a ``dummy'' neural network with input directly attached to the output neurons. The results confirm the findings of \cite{arnold2018deep, chin2020intelligent} that CNNs are superior to fully connected (FC) neural networks for fingerprinting. For example, FC, which was proposed by \cite{chin2020intelligent}, uses a massive $123.6$ million weights compared to \cite{pirnat2022towards} with $3.1$ million weights. However, \cite{pirnat2022towards} achieved nearly $20\%$ better accuracy with less than $2.5\%$ weights. This confirms the findings of \cite{strubell2019energy, strubell2020energy, garcia2019estimation}, where they concluded more weights do not always equate to better model performance. In case of \cite{pirnat2022towards}, more appropriate building blocks (\ie, convolutional layers), optimized kernel size and number of filters led to a huge improvement.

\cite{cerar2021improving} have experimented with different types of building blocks and modifications of the stem, namely CNN4, CNN4R, and CNN4S. The simplest proposed architecture called CNN4 consists of 4 convolutional blocks and a single dense layer. The CNN4R and CNN4S iterations are modifications of CNN4. They achieved an $8\%$ and $13\%$ improvement in accuracy, but at a cost of $103.7\%$ and $207.5\%$ more weights, respectively. Whether this is sustainable is questionable.

As we compare different proposed CNN architecture in Table~\ref{tab:nn:performance}, we see that CNN proposed by \cite{chin2020intelligent} achieved highest accuracy at $0.100\,m$ RMSE. However, while \cite{pirnat2022towards} achieved a slightly worse accuracy at $0.109\,m$ RMSE, it was achieved with only $22.6\%$ of weights of the best performing model.

\begin{table}[htb]
    \ra{1.3}
    \caption{CO$_2$ footprint used in training}
    \label{tab:footprint}
    \centering
    \begin{tabular}{lrrr}
        \toprule
        \bfseries NN & \bfseries carbon footprint & \bfseries FLOPs & \bfseries energy \\
        \midrule
        
        \cite{pirnat2022towards} PirnatEco
        & 10.6\,g\,CO$_2$\,eq.
        & 345\,$\cdot\,10^6$
        & 152\,kJ
        \\
        
        \cite{chin2020intelligent} CNN 
        & 18.3\,g\,CO$_2$\,eq. 
        & 535\,$\cdot\,10^6$
        & 264\,kJ
        \\
        
        \cite{cerar2021improving} CNN4R 
        & 176.9\,g\,CO$_2$\,eq. 
        & 2479\,$\cdot\,10^6$ 
        & 2547\,kJ
        \\
    \bottomrule
    \end{tabular}
\end{table} 

Finally, the best performing models from Table~\ref{tab:nn:performance}  were evaluated in terms of energy consumption and carbon footprint in training. All selected models were trained for 200 epochs. The calculations for selected models are presented in Table~\ref{tab:footprint}, where the first column represents the name of the architecture, second column shows the carbon footprint, in the third we can observe FLOPs of each architecture and in the final column total training energy consumption can be seen. The results show that PirnatEco produces the least amount of carbon footprint compared to the other two models, while achieving similar performance in terms of RMSE. What can also be observed that number of weights does not directly correlate to the carbon footprint, FLOPs and energy consumption, since PirnatEco has approximately 4.4 times less weights compared to CNN from Chin, but has more than half of Chin's carbon footprint. 

Since new generations of mobile networks will heavily rely on mobile phone localization for service quality assurance, it will also impact the energy consumption and carbon footprint of the working model. For example, it is estimated that by the end of 2025 there will be close to 7.4 billion mobile devices in the world and if we used PirnatEco model to predict their location only once, it would result in production of approximately 10 kilograms of CO$_2$.

\section{Summary}

Wireless fingerprinting localization was proven to be an important asset and complement to the GPS for the indoor environment. Most state-of-the-art wireless fingerprinting methods are based on various NN architectures, with CNN being the most popular and successful. Recently, the impact of very powerful AI models raised the concern about their impact on the environment with their power consumption and carbon footprint. In addition, large models require a vast amount of compute resources to obtain, making them difficult to reproduce, which is an issue in research, and exclusive for communities with a large budget and available compute resources. A similar trend can already be observed in the development of wireless localization models.

The chapter examined the energy consumption and carbon footprint of the current state-of-the-art architectures for wireless fingerprinting localization. Both energy consumption and carbon footprint directly correlate to the required floating-point operations (FLOPs) and multiply-accumulate operations (MACs) of the used DL architecture. We presented the methodology for calculating FLOPs for the three most commonly used DL architecture building blocks: fully connected, convolutional, and pooling layers. Then we present an example on how an existing architecture can be adapted for localization to keep good performance and consume significantly less resources.  

To benchmark the state-of-the-art architectures for wireless fingerprinting localization, we used the CTW2019 dataset. In addition, we ran experiments in the same controlled environment where we examined the required number of FLOPs, energy consumption, and their carbon footprint of the model training process.

\section*{Acknowledgements}
The authors would like to acknowledge Anze Pirnat for insightful discussions and the Slovenian Research Agency programme P-0016 for funding this work. 

\bibliography{sn-bibliography}%

\end{document}